\documentclass[twocolumn,prb,color,psfig,superscriptaddress]{revtex4}
\usepackage{graphicx}
\usepackage{epsfig}

\def\beq{\begin{equation}}
\def\eeq{\end{equation}}

\begin{document}

\title{Electron-conformational transformations in nanoscopic  RyR channels govern both the heart's contraction and beating}
\author{A.S.~Moskvin}
\affiliation{Department of Theoretical Physics, Ural State University, 620083 Ekaterinburg,  Russia}
\author{A.M.~Ryvkin}
\affiliation{Institute of Immunology and Physiology, Ural Branch of RAS, 620049, Ekaterinburg,  Russia}
\author{O.E.~Solovyova}
\affiliation{Institute of Immunology and Physiology, Ural Branch of RAS, 620049, Ekaterinburg,  Russia}
\affiliation{Department of Computational Mathematics, Ural State University, 620083 Ekaterinburg, Russia}
\author{V.S.~Markhasin}
\affiliation{Institute of Immunology and Physiology, Ural Branch of RAS, 620049, Ekaterinburg,  Russia}

\date{\today}


\begin{abstract}

 We show that a simple biophysically based electron-conformational model of RyR channel is able to explain and describe on equal footing the oscillatory regime of the heart's cell release unit both in sinoatrial node (pacemaker) cells under normal physiological conditions and in ventricular myocytes under Ca$^{2+}$ SR overload.
 
\end{abstract}

\maketitle

Calcium (Ca$^{2+}$) dynamics is of a principal importance for functioning of different heart's cells from atrial and ventricular cardiomyocites to sinoatrial node cells (SANC) though the former are  responsible for the heart's contraction while the latter for primary heart's  pacemaking, respectively\,\cite{Bers}.   
Cardiac contraction in cardiomyocites is activated by an increase in intracellular calcium concentration (Ca$^{2+}_i$), most of which comes from a specific calcium cistern of sarcoplasmic reticulum (SR). Ca$^{2+}$ is released via the ryanodine receptors (RyR) in response to Ca$^{2+}$ entering the cell via the L-type channels (see Fig.\,1). The cardiac type RyR is the common major Ca$^{2+}$ release channel type in SANC and ventricular myocytes.
It has been experimentally documented in chemically skinned and voltage-clamped SANC, in which effects of voltage-activated sarcolemmal ion currents are excluded, that the isolated SR is capable to spontaneously and rhythmically release Ca$^{2+}$ via RyRs\,\cite{Fabiato,Vinogradova-04}. These spontaneous, rhythmic, local subsarcolemmal Ca$^{2+}$ releases (Ca$^{2+}$  clock), which occur in SANCs,  interact somehow with the classic sarcolemmal voltage oscillator (membrane clock\,\cite{membrane-clock}). 
At present there is a general consensus about the importance of Ca$^{2+}$ oscillator for SANC rate\,\cite{Lipsius}, however, an important discussion still remains whether it is  a dominant or critical factor for cardiac pacemaker cell functioning. Furthermore, the very existence of the intracellular Ca$^{2+}$ clock is not captured by the most part of existing essentially membrane-delimited cardiac pacemaker cell numerical models.

 Recently, Maltsev and Lakatta\,\cite{Maltsev-09} have developed a new numerical SANC model (ML-model) featuring interactions of SR-based Ca$^{2+}$ and membrane clocks to explore novel mechanistic insights into cardiac impulse initiation. 
They started with a well-known simplified model of the cell structure consisting of four compartments: sub-sarcolemmal space (subspace), cytosol, network SR (nSR), and junctional, or luminal SR (jSR) (Fig.\,1).
As in most existing models the authors used an effective medium theory, where Ca$^{2+}$ concentrations in the subspace and in jSR (Ca$_{SS}$ and Ca$_{jSR}$) are main governing parameters that obey standard reaction-diffusion equations, while RyR gating is usually considered in a simplified manner through a dependence of the release on the Ca$^{2+}$ concentrations. The ML-model adopted the formulation of cardiac RyR function developed by Shannon {\it et al.}\,\cite{Shannon-04}) and the Kurata {\it et al.} model\,\cite{Kurata-02} of primary rabbit SANC. Finally the model was formulated in terms of a system of 29 first-order differential equations. 
 The isolated SR can indeed operate as a self-sustained Ca$^{2+}$  oscillator, described by a simple "release-pumping-delay" mechanism: a small spontaneous Ca$^{2+}$  release from jSR to the subspace occurs as the primary or initiating event. When Ca$_{SS}$ increases to a sufficient level, it amplifies the Ca$^{2+}$  release via the mechanism of the Ca$^{2+}$-induced Ca$^{2+}$  release (CICR)\,\cite{Bers}; this relatively strong, secondary Ca$^{2+}$  release simultaneously depletes (i.e., resets) jSR. The released Ca$^{2+}$  is pumped into the nSR. The delay between releases is determined by the Ca$^{2+}$ pumping rate and Ca$^{2+}$  diffusion from the subspace to cytosol and also from nSR to jSR. As Ca$_{jSR}$ slowly increases, RyRs are restituted, and the next release is ultimately initiated, etc.
\begin{figure}[t]
\includegraphics[width=8.5cm,angle=0]{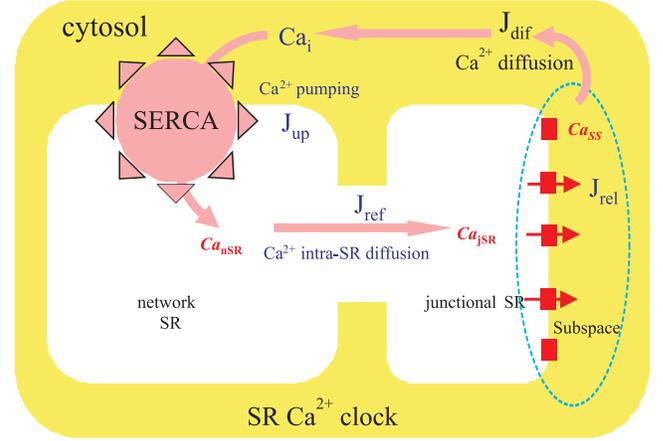}
\caption{Fig.\,1. Schematic illustration of the cell compartments and Ca$^{2+}$ fluxes in the SR Ca$^{2+}$ clock toy model (see below Eqs.(7)-(10). SERCA is a sarco/endoplasmic reticulum Ca$^{2+}$-ATPase that transfers Ca$^{2+}$ from the cytosol  to  the SR.} \label{fig1}
\end{figure}

The ML-model\,\cite{Maltsev-09} of coupled oscillators seems to reproduce basically all recently discovered new behavioral details of cardiac pacemaker cell  function, however, this phenomenological integrative model ignores many important physiological features of  the cardiac cells, in particular, the fine spatiotemporal structure of the Ca$^{2+}$ release. The model of integrated Ca$^{2+}$  dynamics does not describe stochastic, locally propagating Ca$^{2+}$ releases within the subsarcolemmal space. Indeed, RyRs in SANC, as in other cardiomyocites, seem to be arranged in clusters under sarcolemma (Fig.\,1) and thus probably form subsarcolemmal calcium release units (CRUs). In this instance, RyRs release Ca$^{2+}$  into a relatively small volume of subspace where individual jSRs of CRUs approach sarcolemma\,\cite{Bers}. Thus a realistic modeling of  the Ca$^{2+}$ oscillator and  SANC function should include a stochastic mechanism of a local Ca$^{2+}$ release generation by CRUs. Furthermore, main assumption of the ML-model\,\cite{Maltsev-09}, that is the Ca$^{2+}$ released from RyR channels activates the RyR channels as like as the trans-sarcolemmal Ca$^{2+}$ from the L-type channel in a close apposition, seems to be questionable.

 Recently we have applied well-known electron-conformational (EC) model (see, e.g.,Ref.\cite{Rubin}) for a single RyR channel\,\cite{ECM1,ECM2,ECM3} that was shown to capture important features of the individual and cooperative behaviour of RyRs in ventricular myocytes. The EC model of RyR functioning under Ca$^{2+}$ stimuli is based on a biophysical adaptation of the well-known theory of photo-induced structural phase transitions, which has been successfully applied to different solids\,\cite{Koshino}.
Hereafter, in the Letter we will show that EC model of RyR channel is able to explain and describe on equal footing a puzzling spontaneous oscillatory regime of the release unit both in SANC under normal physiological conditions and in ventricular myocytes under Ca$^{2+}$ SR overload.

The ion-activated RyR channel is a giant ($30\times 30$ nm) macromolecular protein
complex comprising 4 subunits of 565 000 Daltons each\,\cite{Bers}.
 As other ion channels it has a great many of internal
electron and conformational degrees of freedom and exhibits
remarkable complexities that need to be considered when developing
realistic models of ion permeation. Nevertheless, until recently
most modelling efforts for RyR channels were focused  on a simple
"hole in the wall" type model with a set of different (open,
closed) states. Our knowledge of molecular mechanisms of RyR
channel functioning is limited; hence we are forced to start with
the most general "physicists'" approach, which is typical for
protein biophysics. Such an approach to the modelling of
biomolecular system implies its simplifying  to bare essentials
with guidance from experimental data. 

Modelling the RyR we start with a simple and a little bit naive picture of
the massive nanoscopic channel like an elastic rubber tube with a
varying cross-section governed by a conformational coordinate $Q$ and
a light "electronic" plug switched due to Ca$^{2+}$-RyR binding/unbinding in the subspace\,\cite{remark}.
This electronic plug interacts with the conformational coordinate and acts as
a trigger to stimulate its change and related channel
cross-section/conductivity. 
\begin{figure}[t]
\includegraphics[width=8.5cm,angle=0]{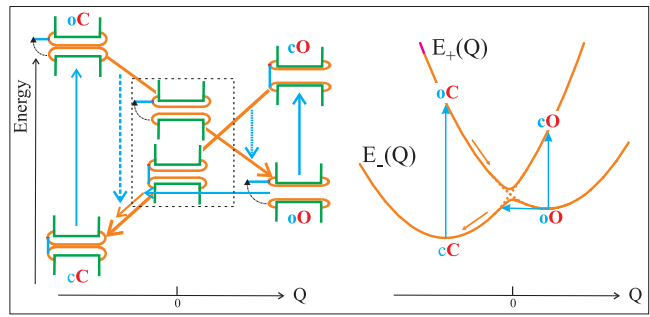}
\caption{Fig.\,2. Left panel: Naive "tube-and-plug" model  of RyR channel in a closed ground state. Small letters $c,o$ are used for  electronically closed and open states, respectively, and capital letters $C,O$ for conformationally closed and fully open states, respectively. Right panel: Corresponding adiabatic potentials in EC-model of RyR.  Vertical arrows point to Ca$^{2+}$-induced Franck-Condon (FC)  electronic transitions, horizontal arrow points to a non-FC tunneling transition, downhill arrows point to a conformational dynamics. } \label{fig2}
\end{figure}
In other words, we reduce a large variety of RyR degrees of
freedom to only two: a fast and a slow one, conventionally termed as
electronic and conformational one, respectively. Both degrees of
freedom are implied to be coupled to realize an
EC transformation that is the electronic
control of the slow conformational motion. Bearing in mind the main
function of RyR channels, we assume only two actual electronic RyR
states: "open" and "closed", and a single conformational degree of
freedom, $Q$, described by a classical continuous variable. Figure 2 illustrates the model with a set of representative states of the system.

Change in the electronic and conformational states regulate the main RyR channel function, i.e. determines whether the channel is "open"  and permeable for Ca$^{2+}$ ions or "closed" and impermeable to ions. Hereafter we assume that the conformational variable Q specifies the RyR channel "cross-section" or, more precisely, a permeability for Ca$^{2+}$, while the dichotomic electronic variable determines its opening and closure. This allows us to describe the Ca$^{2+}$ flux through the RyR as follows:
\begin{eqnarray}
J_{RyR}=D(Q)(Ca_{jSR}-Ca_{SS}),
\label{J}
\end{eqnarray}
if the channel is electronically open, and $J_{RyR}$ = 0, if it is closed. Here,
the permeability coefficient $D(Q)$ reflects the ease with which
Ca$^{2+}$ passes through an open RyR. Its functional dependence on
the conformational coordinate should be one of the essential model
assumptions. $D(Q)$ is assumed to be an increasing function of the conformational coordinate,
varying from zero or small leakage value to some saturated value
$D_0\,\, (0<D(Q)<D_0)$, when $Q$ varies from large negative to
large positive magnitudes, passing through some subconductive
state at $Q = 0$.
As a simplest limiting case, we may
consider the step-like dependence $D(Q < 0) = 0, D(Q \geq 0) =
D_0$. 

Hereafter, we shall assume a simple
harmonic approximation for the conformational energy and use a 
 Hookes harmonic law for elastic potential energy:
$E=\frac{1}{2}KQ^2$, where $K$ is the effective "elastic" constant
and $Q = 0$ relates to a base state with "unstrained tube" and a
bare cross-section. It is worth noting that namely the EC model introduces the energy to be an important  factor of the RyR functioning.

As a starting point of the EC model algebra we introduce a simple
effective  Hamiltonian for a single RyR
channel as follows\,\cite{ECM1,ECM2,ECM3}
 \begin{equation}
H_{s}= -\Delta{\hat s}_{z} -h{\hat s}_{x}- pQ +\frac{K}{2}Q^2 + aQ{\hat s}_{z}\, , \label{H}
\end{equation}
where $s_z$ and
 $s_x$ are well-known Pauli matrices, and the first term describes the bare energy
splitting of "up" and "down" (electronically "open" and "closed") states with an
energy gap $\Delta$, while the second term describes their mixing. It is worth noting that given $h = 0$ we arrive at
a classical approach with a dichotomic electronic variable. The
third and fourth terms in (\ref{H}) describe the linear and
quadratic contributions to the conformational energy. Here, the
linear term formally corresponds to the energy of an external
conformational stress, described by an effective stress parameter $p$.
The last term describes the EC interaction with the coupling parameter $a$. Hereafter we make use
of the dimensionless conformational variable $Q$; therefore all of
the model parameters ($\Delta$, $h, p, K, a$) are assigned energy units. 
 Two eigenvalues of our Hamiltonian
\begin{equation}
E_{\pm}(Q)=\frac{K}{2}Q^2 - pQ \pm\frac{1}{2}\left[(\Delta-aQ)^2 + h^2
\right]^{1/2}
\end{equation}
define two branches of the adiabatic, or conformational potential (CP), attributed to electronically closed ($E_-$) and
electronically open ($E_+$) states of the RyR, respectively (see Fig.\,2).

The classical dynamics of the conformational coordinate we assume to
obey a conventional Langevin equation of motion
\begin{equation}
M\ddot{Q}=-\frac{\partial}{\partial Q}E(Q)-M\,\Gamma \dot{Q}+R(t),
\label{equa}
\end{equation}
where first term describes a total systematic  conformational
force with $M$ being an effective RyR mass (below $M$ let to be unity), $\Gamma $ is an effective dimensionless friction damping constant (relaxation rate), and $R$ is the thermal fluctuation
force (Gaussian-Markovian noise). The last two terms reflect the
coupling to an external environment of channels. 

 The thermo-activated transitions  are caused by the thermal
fluctuating force and free from the Franck-Condon principle.  As a
result of thermal transitions the whole system will finally relax
to the thermal equilibrium. The temperature plays an important
role in overcoming the energy barriers. 
In fact we deal with a hybrid "over" (thermal
activation) and "through" (quantum tunneling) barrier transfer
transitions (reactions). Quantum tunneling is often
addressed to specify the low-temperature limit of the barrier
transfer transition probability that points to the possible way to
uncover non-classical behaviour for the RyR channel. 
We assumed the resonant  quantum tunneling takes place between two branches of conformational potential starting within a tunneling zone $\delta Q=\epsilon$ centered at the CP minimum with  the probability  obeyed the effective Gamov law   
\begin{equation}
	P_{tun}=P_{0}e^{-A_{tun}\Delta Q\sqrt{\Delta E}} \, ,
\end{equation}
where $\Delta Q$ is the width,  $\Delta E$ is the height of the energy barrier, or the energy separation between the tunneling points and the point of the two branch intersection (see Fig.\,2), and $P_{0}$ effective tunneling attempt frequency.

At present there is no sufficient understanding of the
mechanisms that regulate local Ca$^{2+}$ signaling in heart's cell
despite persistent efforts to discriminate between the cytosolic
and luminal Ca$^{2+}$ activation hypotheses. 
Calcium enters the subspace by two pathways: across the sarcolemma via L-type channels and from the SR via RyR channels. 
Activation of RyRs by  Ca$_{jSR}$ has been attributed to either Ca$^{2+}$ feedthrough to high-affinity cytoplasmic  Ca$^{2+}$ activation sites  or to Ca$^{2+}$ regulatory sites on the luminal side of the RyR. However, most of experimental observations on cardiac RyRs more difficult to reconcile with the Ca$^{2+}$ feedthrough effect\,\cite{Gyorke}. 
A luminal Ca$^{2+}$ sensor appears to continuously regulate the functional activity of the SR Ca$^{2+}$ stores by linking SR Ca$^{2+}$ content to the activity of the RyRs\,\cite{Gyorke}.
However, in contrast with purely electronic effect of Ca$_{SS}$, the effect of relatively slowly varying Ca$_{jSR}$ on the RyR channels is likely to be purely "mechanical" one, through the respective conformational strain applied to RyR channels. The effect can be naturally incorporated into EC model, if we assume the strain parameter $p$ in the RyR Hamiltonian to be a function of Ca$_{jSR}$ or the jSR-to-subspace Ca$^{2+}$ gradient.  
We assume $p$ to rise with the luminal Ca$^{2+}$ concentration in accordance with the Hill curve\,\cite{ECM2}:
 \begin{equation}
    p=2\frac{[Ca_{jSR}]^n}{[Ca_{jSR}]^n+K_{Ca}^n}-1; \, -1\leq p \leq +1 \, ,
\end{equation}
where $K_{Ca}$  is the half maximal value, $n$  is a Hill coefficient. Rise of $p$ in the interval (-1,+1) leads to a crucial modification of CP from that of stabilizing closed RyR state to that of stabilizing open RyR state. It should be noted that in terms of the EC model one might introduce a critical SR load that specifies a critical effective strain $p_{cr}=\left(\frac{a}{2}-\frac{K}{a}\Delta\right)$ when  the minimum of the CP$_c$ branch for the electronically closed RyR state crosses CP$_o$ branch thus destroying the RyR bistability conditions and making RyRs stay close to their activation threshold where their cC$\rightarrow$oO activation can presumably be easily provoked.

Cooperative dynamics of the RyR clusters in CRU  has been studied in a series of model simulations for 11$\times$11 square RyR lattice in our prior paper\,\cite{ECM3}. We simulated both "in vitro" dynamics of RyR lattice when the calcium surrounding was assumed to be simply  as a source of effective external fields, such as effective strain,  and "in vivo" situation when the RyR lattice dynamics is incorporated into the whole cell calcium dynamics (EC-CRU model). 
In such a case the EC dynamics of RyR lattice is assumed to specify the number of open channels $N_{open}$ which in its turn specifies the release flux from the SR to subspace through RyR channels: $J_{rel}=k_{rel}N_{open}(Ca_{jSR}-Ca_{SS})$, where $k_{rel}$ is a single RyR channel release velocity coefficient. 
The system demonstrated four different modes of
behaviour, depending on the SR load\,\cite{ECM3}: inactivation mode I at a rather small SR load (heavily underloaded SR);
single channel activated mode II (Ca synapse, or quark mode)  at underloaded SR;
domino-like firing-termination mode III with high degree of cooperativity (cluster bomb mode) at optimally loaded SR.
The cluster bomb mode is characterized by a  step-by-step opening of the neighboring RyRs with formation of a cluster composed of  up to 40\% open RyRs. The process results in an effective high-gain Ca$^{2+}$
release from the SR. However, the decrease in SR
load leads to a lowering of the effective strain, accompanied by
the shift of the system to preferable channel's closure. This
negative feedback effect results, first, in a slowing down of the
nucleation process and, second, in an evolution of the inverse
domino-like effect with full collapse of the cluster of open RyRs
and termination of Ca$^{2+}$ release. 
Puzzlingly, the CRU simulation\,\cite{ECM3} showed that 
SR overload can result in the excitation of the RyR lattice
auto-oscillations when the domino-like opening of
a cluster of RyR channels resulted in effective Ca$^{2+}$ release that,
in turn, caused a lowering of the effective strain and
simultaneous closure of RyRs before the SR load
started rising. However, until Ca$_{jSR}$ approached the initial value,
close to a critical concentration all channels re-opened simultaneously,
and Ca$^{2+}$ release repeated spontaneously. This behaviour is
repetitive, i.e. the system turned out to behave in an
auto-oscillatory mode IV with a spontaneous SR Ca$^{2+}$ release.
It should be emphasized that in contrast with the ML model, the auto-oscillation mode IV occurs as a result of the Ca$_{jSR}$-dependent shape of the CP and a purely conformational $cC\leftrightarrow oO$ transformation without any L-type channel activity and Ca$^{2+}$-induced electronic $c\leftrightarrow o$ transitions. 
To the best of our knowledge it was a first quantitative model for the CRU oscillatory mode.
\begin{figure*}[t]
\includegraphics[width=17.0cm,angle=0]{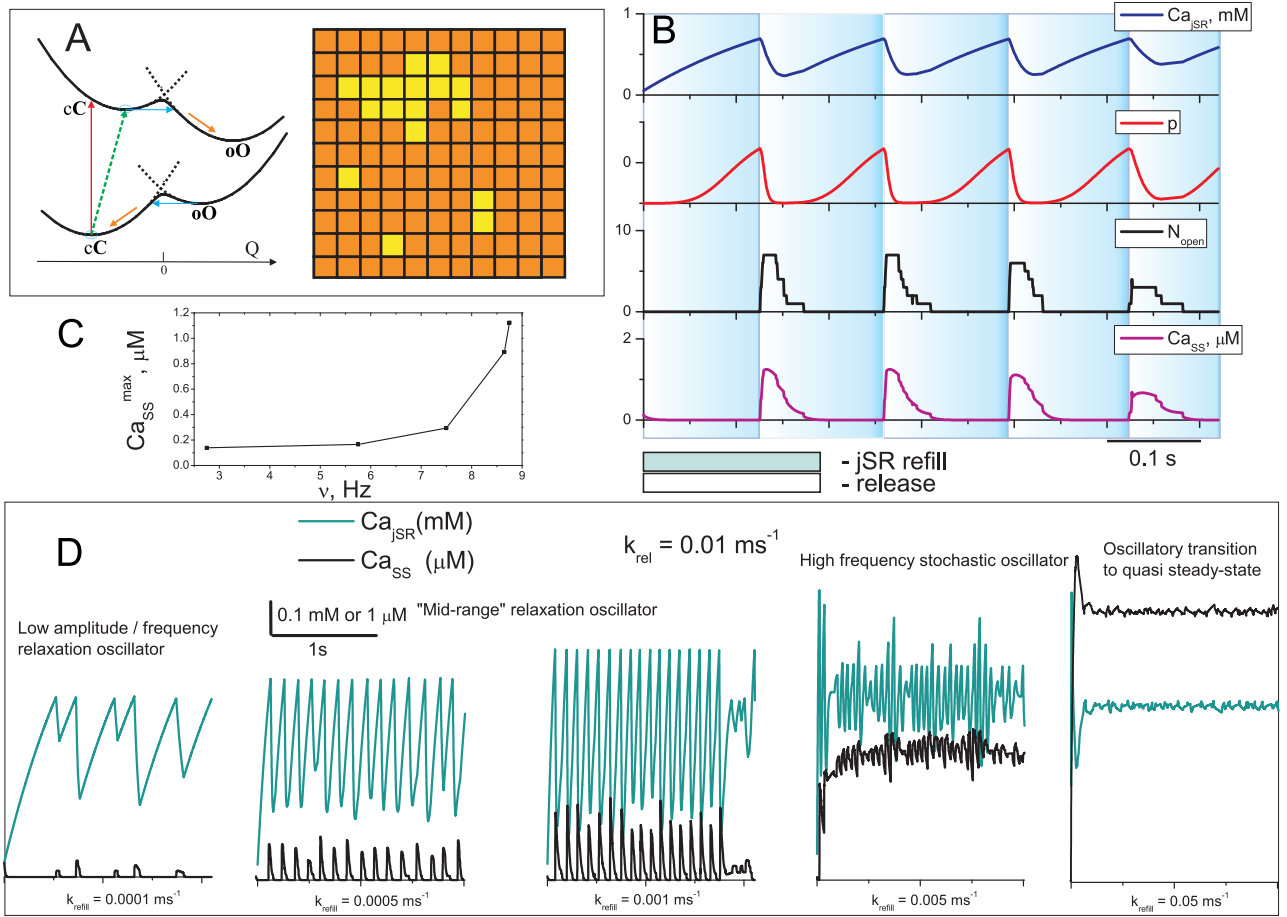}
\caption{Fig.\,3. A: Illustration of a repetitive shaping of the RyR's CP branches during refill-release mode and a flash pattern of open and closed RyRs in our 11$\times$11 cluster. B: Model simulations illustrating that CRU can operate as a self-sustained Ca$^{2+}$  oscillator. C: SR Ca$^{2+}$ oscillator operates following a Bowditch-like pattern: "the faster rate, the stronger release". Bottom panel D: Ca$_{SS}$ and Ca$_{jSR}$ courses given different refill rate constant $k_{ref}=10^{-4}\div 5\cdot 10^{-2}\,ms^{-1}$ at $k_{rel}=0.01\,ms^{-1}$, $k_{dif}=5\,ms^{-1}$, $P_{up}=1\,\mu M/ms$. We make use of the EC model parameters as follows: $\Delta =0,\,K=12,\,a=5\,,\Gamma =7,\,P_{tun}^0 = 0.1,\,A_{tun} = 20,\,\epsilon= 0.001$.} \label{fig3}
\end{figure*}

Obviously the abnormal Ca$^{2+}$ automaticity in SR overloaded ventricular myocytes suggests  the normal pacemaker SANC activity may have some similar features and actually proceed with elevated Ca$^{2+}$ level. Indeed, Vinogradova {\it et al.}\,\cite{Vinogradova-04} found a minimum diastolic Ca$^{2+}$ level in rabbit SANC of $\sim$\,200\,nM, that is twice the diastolic Ca$^{2+}$ level in resting ventricular myocytes.

Hereafter in the paper we address a more realistic EC-CRU model relevant for the SANCs which incorporates two main features of the Ca$^{2+}$ machinery, that is a local and stochastic character of the Ca$^{2+}$ release. As in our prior paper\,\cite{ECM3} we address the CRU model with 11$\times$11 square lattice of RyRs, however, for simplicity we neglect any RyR-RyR coupling. At variance with \cite{ECM3} each RyR channel was described by a diabatic CP with non-FC tunneling transition between two CP branches. 
The calcium fluxes in a simplified cell model (Fig.\,1) were assumed to obey a standard system of four differential equations\,\cite{Sobie}:
\begin{equation}
\frac{dCa_{SS}}{dt}=(k^{jSR}_{SS}J_{rel}-J_{diff});\,
\end{equation}
\begin{equation}
\frac{dCa_{i}}{dt}=(k_i^{SS}J_{diff}-J_{up});
\end{equation}
\begin{equation}
\frac{dCa_{nSR}}{dt}=(k^{i}_{nSR}J_{up}-J_{ref});\,
\end{equation}
\begin{equation}
\frac{dCa_{jSR}}{dt}=(k^{nSR}_{jSR}J_{ref}-J_{rel})\, ,
\end{equation}
where $J_{ref}=k_{ref}\left(Ca_{nSR}-Ca_{jSR}\right)$, $J_{rel}=N_{open}k_{rel}\left(Ca_{jSR}-Ca_{SS}\right)$, $J_{diff}=k_{diff}\left(Ca_{SS}-Ca_{i}\right)$ are diffusion fluxes 
between the nSR  and the jSR, between the jSR and subspace, between subspace and cytosol, respectively, $J_{up}=\frac{P_{up}\,Ca_i}{K_{up}+Ca_i}$ is  uptake flux;  
$k^{\alpha}_{\beta}$ are volume ratio constants between $\alpha$ and $\beta$ cell compartments ($k_{SS}^{jSR}=0.12,\,k^{SS}_{i}=0.022,\,k_{nSR}^{i}=40,\,k^{nSR}_{jSR}=9.7$).
 It should be noted that all the parameters were chosen rather as typical for integrative cell models  than for a
single CRU model. In other words, all the cell RyRs we consider to form a system of identical CRU's functioning in concert.
We set initial conditions as follows: Ca$_i=0.1\,\mu M$, Ca$_{SS}=0.5\,\mu M$, Ca$_{jSR}=0.05\,mM$, Ca$_{nSR}=1.5\,mM$, for other Ca$^{2+}$ dynamics parameters:$K_{up}=0.0006\,mM;\,k_{diff}=5\,ms^{-1}$.
 As in the ML model the Ca$_{SS}$ and Ca$_{jSR}$ time courses strongly depend on the refill and release rate constants, that are expected both to change from their original values of ventricular myocytes and likely mediate the regulation of SR Ca$^{2+}$  clock ticking speed\,\cite{Maltsev-09}. Main results of numerical simulations are presented in Fig.\,3.
 
 As in ML model, three types of the steady-state were found: 1) steady rhythmic oscillations; 2) no oscillations or damped oscillations; and 3) chaotic oscillations.
 The highest rates are reached when the oscillator approaches to dynamic equilibrium (steady release, Fig.\,3d, right part). The lowest rates are reached when the oscillator approaches to static equilibrium.
Static equilibrium occurs when either the release rate is too small or Ca$^{2+}$ pumping rate is too fast, i.e., jSR becomes highly loaded.  
  Fig.\,3 distinctly shows that CRU can operate as a self-sustained Ca$^{2+}$  oscillator, however, the triggering of the Ca$^{2+}$ release is determined by the bringing  Ca$_{jSR}$ near the critical concentration, or activation threshold rather than by a small initial spontaneous Ca$^{2+}$ release from jSR as in the ML-model\,\cite{Maltsev-09}. Maximal Ca$_{jSR}$ appears to slightly rise with $k_{ref}$, while its oscillation amplitude strongly depends on $k_{ref}$, however, in an absolutely different way than in ML-model. Indeed, for a rather wide range of $k_{ref}$ the Ca$_{jSR}$ oscillation amplitude rises with the refill rate, however, without full depletion of the jSR as it occurs in ML-model. Furthermore,  in contrast with the ML-model the EC-CRU model reveals the Bowditch-like behavior: "the faster rate, the stronger release". Indeed, within oscillatory regime both the frequency and amplitude of Ca$_{SS}$ do increase with $k_{ref}$.  A detailed analysis shows  this can be explained as a result of some $retardation$ effect due to a slow conformational dynamics near the CP minima; optimal condition for the inter-branch tunneling and Ca$^{2+}$ release are realized at higher $k_{ref}$ and consequently, at higher Ca$_{jSR}$. Indeed, a high-speed increase of the Ca$_{jSR}$ results in a rapid rise of effective stress $p$ with a shift of the CP branches with its minima up and right (see Fig.\,3). Given rather slow conformational dynamics the CP shift leads to a tunneling retardation effect as the conformational coordinate $Q$ needs a time to reach the new tunneling zone $\epsilon$ in the vicinity of new CP minimum. In other words, the EC system is not in time for adjusting to the rapid shift of the CP branches. However, the higher Ca$_{jSR}$ the slower the CP shift, that allows conformational coordinate $Q$ to run tunneling zone down, and provide the optimal condition for the inter-branch tunneling and Ca$^{2+}$ release. Despite the tunneling retardation, the bigger $k_{ref}$  the shorter time  to reach the critical Ca$_{jSR}$ level.
  
Just the opposite behavior, "the higher the rate, the lower the amplitude," represents the major limitation of the isolated Ca$^{2+}$ oscillator in frames of ML-model: it is unable to generate high-amplitude oscillations at higher rates. However, the authors\,\cite{Maltsev-09} have shown that the interactions of the SR and sarcolemma clocks  overcomes the limitation of the isolated SR Ca$^{2+}$ clock, i.e., the oscillation amplitude of the full system Ca$^{2+}$ clock raises as the oscillation rate increases, i.e., the Bowditch phenomenon  restores. At the same time 
the "Bowditch-like" behavior (the faster the stronger) for the both Ca$_{SS}$ and Ca$_{jSR}$ appears to be a distinct feature of the EC-CRU model.
 In other words, it seems the SR Ca$^{2+}$  oscillator can serve as a dominant source of persistent heart's beating. Thus the interrelation between two types of cell oscillators needs in a futher examination.

In summary, at variance with the integrative model by Maltsev and Lakatta simple biophysically based EC model of RyR channel describes stochastic local Ca$^{2+}$ releases within the subsarcolemmal space as a result of conformational transformations followed by a tunneling between two CP branches.
The model is able to explain and describe the spontaneous oscillatory regime of the CRU both in pacemaker cells under normal physiological conditions and in ventricular myocytes under Ca$^{2+}$ SR overload including its subtle features such as the amplitude and frequency fluctuations. At variance with the integrative ML-model, the oscillation amplitude of the intracellular  Ca$^{2+}$ clock raises as the oscillation rate increases, thus providing  the Bowditch law functioning of the pacemaker cell without any membrane clock assistance.
Despite the EC model is intentionally
simplistic, it offers novel insight into potential mechanisms governing by the Ca$^{2+}$ fluxes and may thus
provide a starting point for further exploration of physical
principles guiding cardiac cell functioning {\it in vitro} and {\it in vivo}.

We acknowledge the support by Ural Branch of RAS under grant No.09-M-14-2001.


\begin{thebibliography}{99}
\bibitem{remark}
At variance with the ion-activated RyR channel the potential-activated channels can be described by a single conformational coordinate governed by the electric field.
\end{thebibliography}
\end{document}